# Gate controlled photocurrent generation mechanisms in high-gain In$_2$Se$_3$ phototransistors

J. O. Island*, S. I. Blanter*, M. Buscema, H.S.J. van der Zant, A. Castellanos-Gomez
Kavli Institute of Nanoscience, Delft University of Technology, Lorentzweg 1, 2628 CJ Delft, The Netherlands.
*Equal contribution
E-mail: j.o.island@tudelft.nl,  a.castellanosgomez@tudelft.nl

**Abstract:**

Photocurrent in photodetectors incorporating van der Waals materials is typically produced by a combination of photocurrent generation mechanisms that occur simultaneously during operation. Because of this, response times in these devices often yield to slower, high gain processes which cannot be turned off. Here we report on photodetectors incorporating the layered material In$_2$Se$_3$, which allow complete modulation of a high gain, photogating mechanism in the ON state in favor of fast photoconduction in the OFF state. While photoconduction is largely gate independent, photocurrent from the photogating effect is strongly modulated through application of a back gate voltage. By varying the back gate, we demonstrate control over the dominant mechanism responsible for photocurrent generation. Furthermore, due to the strong photogating effect, these direct-band gap, multi-layer phototransistors produce ultra-high gains of $(9.8 \pm 2.5) \cdot 10^4$ A/W and inferred detectivities of $(3.3 \pm 0.8) \cdot 10^{13}$ Jones, putting In$_2$Se$_3$ amongst the most sensitive 2D materials for photodetection studied to date.

**Introduction:**

Photodetection incorporating layered, van der Waals materials is a rapidly expanding field.[1] Strong light-matter interaction, large surface-to-volume ratio, transparency and flexibility all make two-dimensional (2D), layered materials strong candidates as next generation photodetectors.[2-13] Photodetectors using these materials often display high gains (MoS$_2$, TiS$_3$, GaTe) and even fast response (graphene) but the underlying photocurrent generation mechanisms are typically a combination of several processes. MoS$_2$ photodetectors, for example, are known to display signatures of the photoconductive (PC), photoelectric (PE), photo-thermoelectric (PTE), and photovoltaic effects (PV).[14-16] Depending on operating conditions, all or some of these mechanisms are playing a role in the total photoresponse making it difficult to characterize and control a single property of the photodetector such as gain or response time. The ever-growing family of layered chalcogenides possess other materials, however, that could be more suitable for photodetection and allow control over the inherent mechanisms. Indium selenide (In$_2$Se$_3$) for example, having a direct bandgap of 1.45 eV,[17] has already shown promise as a high performance, two-terminal photodetector[18, 19] and as a solar cell material[20, 21] but a comprehensive study of the photocurrent generation mechanisms in gate-controlled In$_2$Se$_3$ photodetectors is lacking.

Here, we report on the fabrication and characterization of multi-layer In$_2$Se$_3$ phototransistors and realize control over the dominate photocurrent generation mechanism allowing full modulation of the gain in favor of response time. Under illumination and low bias ($V_{ds}$ = 50 mV), the devices produce ultra-high responsivities up to ~10$^5$ A/W, outperforming the majority of photodetectors based on other 2D materials such as graphene, MoS$_2$, and even highly sensitive multi-layer GaTe flakes (responsivities reaching 10$^4$ A/W).[3, 11, 22, 23] We demonstrate control over the dominant photoresponse by application of a back gate, tuning the optical gain from a fully linear dependence on illumination power (fast photoconduction) to a sublinear (high-gain photogating) dependence. Finally, we measure the spectral response and discuss the potential of In$_2$Se$_3$ in photodetector applications by calculating the inferred detectivity. We find a broad spectrum response across the visible wavelengths and calculate inferred detectivites up to $3 \cdot 10^{13}$ Jones making In$_2$Se$_3$ very attractive for optoelectronic applications.

**Results:**

Single layers of In$_2$Se$_3$ are formed by a five-layer sandwich of covalently bonded indium and selenium atoms, shown in Figure 1a, which are stacked on top of each other and held together by van der Waals forces. Currently, five phases of the crystal are known to exist (α, β, γ, δ, and κ) which are formed by different arrangements of indium and selenium atoms within in a single layer.[24, 25] For this study, we isolate multi-layer α-In$_2$Se$_3$ (see Supporting Information for Raman and optical absorption characteristics which identifies the unique alpha phase) flakes and transfer them to a Si/SiO$_2$ substrate to fabricate field-effect transistors (FET) (see methods for fabrication details). Figure 1b shows an AFM image of a fabricated device. The channel length (the distance between two consecutive pairs of leads) varies from 1 μm to 3 μm. The height of the contacted part of the flake is 14 ± 5 nm (see Supporting Information for an AFM height scan), corresponding to about 15 ± 5 layers of In$_2$Se$_3$.[26] We characterized the devices in dark and under illumination and all measurements were performed in a probe station at room temperature and in vacuum ($10^{-5}$ mbar). Three separate FET devices were measured in detail with comparable characteristics and we present one in the main text and another in the supporting information.

In dark conditions, Figure 1c shows the transfer curve ($I$-$V_g$) at 100 mV voltage bias ($V$) and Figure 1d shows current-voltage ($I$-$V$) curves as function of gate voltage ($V_g$). A typical n-type behavior is observed; the current magnitude is the highest (~ 1 µA for $V$ = 50 mV) with positive gate voltages (electron accumulation) and decreases to below 1 nA for $V_g$ < -20 $V$. From the transfer curve in figure 1c, we determine an ON/OFF ratio of $10^5$. Using the transmission line model (TLM)[27, 28], we estimate the field-effect mobility, $\mu$, using the following formula:

$$R_{tot} = R_c + \frac{L}{WC_{ox}\mu(V_g-V_t)} \quad (1)$$

where R$_{tot}$ is the total measured resistance, R$_c$ is the contact resistance (from a linear fit of the resistances versus channel length), $L$ and $W$ are the channel length and width, respectively and $C_{ox}$ the parallel plate capacitance to the gate. The length of the channel is the distance between adjacent contacts and the width is the overlap distance of the electrode and flake. In an attempt to only contact the thinnest most homogenous portion of the flake, the contacts on the main text device overlap the flake only partially. We have included an error of 25% in the width determination to account for current fringing outside the channel defined by the electrode overlap (see Methods). At a gate voltage of 40 V we estimate mobilities up to 30 ± 8 cm$^2$/Vs.

We now turn to the optoelectronic characterization and mechanisms responsible for photocarrier generation in these In$_2$Se$_3$ photodetectors. Figure 2a shows the device transfer curves ($V$ = 50 mV) taken at increasing laser powers ($\lambda$ = 640 nm). Higher powers are seen to increase the drain current of the photodetector over the full range of the gate voltage. The threshold voltage ($V_t$) is extrapolated from the linear portion of the transfer curves (dotted line in Figure 2a). The inset of Figure 2a shows the shift in threshold voltage extracted for each transfer curve. The threshold voltage sharply decreases for low powers and saturates at higher powers. In Figure 2b we plot the photocurrent ($I_{ph} = I_{illumination} - I_{dark}$) as a function of the effective incident laser power ($P_{eff} = P_{in} \cdot A_{device}/A_{laserspot}$) for different gate voltages ranging from -40 V to +40 V. From the photocurrent we calculate the responsivity ($R = I_{ph}/P_{eff}$), shown in Figure 2c as a function of gate voltage, which reaches $(9.8 \pm 2.5) \cdot 10^4$ A/W at $V_g$ = 30 V at the relatively low bias voltage of 50 mV. To our knowledge this is the highest reported responsivity for a photodetector incorporating a two-dimensional, van der Waals material.[1] In Table 1 we show a comparison of other reported responsivities and response times for similar multi-layered photodetectors.

The main photocurrent generation mechanisms in layered material phototransistors are the photoconductive effect, photovoltaic effect, and thermal mechanisms: photo-thermoelectric effect and bolometric effect.[1] Here, we employ a large laser spot (200 μm diameter) and therefore we do not expect sharp thermal gradients that would give rise to any appreciable contribution from thermal mechanisms to the total photocurrent. At the highest applied power density (~1μW/μm$^2$), we estimate temperature gradients less than 0.1 K[14] which, given a Seebeck coefficient of 200 μV/K[29], translates to a negligible contribution from the photo-thermoelectric effect of 20 μV given that the applied bias voltage is 50 mV.

Photocurrent from the photoconductive effect arises from an increase in drain current from photogenerated carriers, $I_{PC} = q\mu nEWD$, where $\mu$ is the carrier mobility, $n$ is the excess carrier density, $E$ is the electric field in the channel, $W$ is the channel width, and $D$ is the absorption depth.[30] A signature of the photoconductive effect is a linear increase in measured photocurrent with incident laser power ($I_{PC} \propto P_{eff}$).[30, 31] Photocurrent from the photovoltaic effect, on the other hand, arises from a shift in a transistor's threshold voltage giving rise to an increased drain current, $I_{PV} = g\Delta V_t$, where $g$ is the transconductance, and $\Delta V_t$ is the shift in threshold voltage, and generally increases sublinearly with incident laser power ($I_{PC} \propto P_{eff}^\alpha$, for $\alpha < 1$).[31, 32] In layered material phototransistors, owing to the large surface to volume ratio, the photovoltaic effect manifests itself as photogating which is mediated by long-lived states from surface and interface traps.[11, 16, 33] In the case of In$_2$Se$_3$, traps are most likely a result of the natural surface oxide which forms because of vacancies in the basal plane layers of the material.[17] Surface oxides have been shown to be the cause for hole trapping giving rise to photogating effects in ZnO and GaN nanowire photodetectors.[34, 35] We interpret then the measured shift in threshold voltage, plotted in the inset of Figure 2a, as the dominant photogating mechanism allowing ultra-high gain in these In$_2$Se$_3$ phototransistors. Presumably, photogenerated holes are trapped in long-lived surface oxide and interface states and effectively gate the flake leading to increased electron carrier concentration. Charge traps are also evident from the large hysteresis in forward and back gate sweeps which can be found in the Supporting Information. At higher powers, these traps become saturated and the shift in $V_t$ reduces (also apparent in the inset of Figure 2a at higher powers). Trap saturation is again evident from the dependence of the responsivity on laser power (Figure 2c). Higher incident laser powers result in saturation of the trap states responsible for high gain and lead to monotonically decreasing gain.[11, 33, 36, 37]

Given the power dependence of the photogenerated current from the photoconductive and photogating effects, we capture the dominant mechanism through a simple power-law, $I_{PC} \propto P_{eff}^{\alpha}$, where we extract α from a linear fit to the log-plotted data in Figure 2b. We plot the extracted α as a function of gate voltage in Figure 2d. It can be seen that α ranges continuously from ≈ 1 (photoconduction) in the OFF state at -40 V to < 1 (photogating) across the ON state (see Supporting Information for photocurrent data of contacts 1-2 and 2-3). While the photoconductive effect is largely gate independent, the photogating effect is directly dependent on the transconductance and shows a strong modulation with back gate voltage. Thus, by varying the gate voltage, we tune the dominant photocurrent contribution from photogating to photoconduction. An estimate of the contribution from photogating in the OFF state confirms this. At a gate voltage of -40 V to -35 V the transconductance is ≈ 5x10$^{-14}$ A/V (from a linear fit) and at a power of 1 mW the threshold voltage shift is roughly 20 V. This gives an estimate ≈ 1 pA for the photogating contribution to the total photocurrent which is orders of magnitude less than the photocurrent measured in the linear regime at $V_g$ = -40V (100 nA at 1 mW).

Further insight on these mechanisms is obtained by measuring the time response of the photocurrent in the ON and OFF state by modulating the intensity of the laser using a mechanical chopper. Figure 3 shows the time response of the device obtained by illuminating the sample at a wavelength of 640 nm ($P_{eff}$ = 960 nW) at $V_g$ = 0V ( > $V_t$ at the effective power). We distinguish between two different time responses in the ON state. Figure 3a shows the measured drain current for a single on/off cycle of the photodetector. A slow response of the detector is seen which we attribute to the slow traps responsible for high gain. The fall time (taken between 90% and 10% of the total drain current) is ≈ 9 s and comparable with fall times as a result of slow oxide traps in ZnO and GaN nanowire photodetectors.[34, 35] On top of this strong, slow response in the ON state, we measure a weaker, fast response by modulating the incident laser at a frequency of 10 Hz. The inset of figure 3a shows the on/off characteristics of this signal which has a fall time of ≈ 30 ms. In the OFF state ($V_g$ = -40 V ) the high gain, slow response is absent but the faster response is still present which we measure again at a chopper frequency of 10 Hz (see Figure 3b for a single, averaged on/off cycle). We attribute the fast response to the intrinsic photoconduction mechanism which is present in the ON and OFF states of operation. Photoconduction, which does not rely on slow oxide traps, allows faster response of the photodetector. The ratio of the gains for the two signals ($R_{slow}$/$R_{fast}$ ~ 10$^4$) shows again the large gain coming from the photogating effect as the dominant mechanism found in the steady state characterization above (Figure 2). By varying the backgate, the slow photogating contribution to the total photocurrent can be tuned below that of the photoconductive contribution to allow orders of magnitude faster response times.. This control allows smooth variation of the gain in favor of fast response which is not commonly found in layered material photodetectors where two or more photocurrent generation processes are competing.

Above, all measurements have been made at a wavelength of 640 nm. To ascertain further the sources of high photogain, we measure the spectral response of the detector across wavelengths of 405 nm to 940 nm. Figure 4a shows the measured drain current as function of gate voltage for different wavelengths and in the dark state. One can observe a clear photoresponse for a broad range of wavelengths. In has been shown that the intrinsic defects and native oxide that grows at the surface of the In$_2$Se$_3$ flakes in ambient conditions act as efficient energy converters of incident light which supports this broad response.[17] In Figure 4b we plot the photocurrent as a function of wavelength. A peak is discernable in the photocurrent between 532 nm and 808 nm for all gate voltages corresponding to the band gap energy of the native oxide layer (2.18 eV, 569 nm).[17] In this energy range, the oxide layer efficiently absorbs light which enhances the photogain.

With these experimental observations in mind, we sketch a qualitative picture for the origin of ultra-high gain in these In$_2$Se$_3$ photodetectors. The oxide layer increases the efficiency of light absorption at higher energies. Aditionally, photogenerated holes are trapped in long lived states at the oxide interface which gate the flake and further increase the drain current. Furthermore, while the dichalcogenides show a direct band gap only in single-layer form, In$_2$Se$_3$ has a direct band gap in bulk. The total absorbance (I) of a photodetector is proportional to the material's absorption constant and thickness, $I = I_0[1 − \exp(−\alpha d)]$, where $\alpha$ is the material's absorption constant and $d$ is the thickness. We expect that our thicker devices benefit from higher absorption when compared with single layer dichalcogenides.

Finally, envisioning the use of In$_2$Se$_3$ as a photodetector for applications we calculate the inferred detectivity ($D^*$). The detectivity is a common figure-of-merit which makes possible a direct comparison between photodetectors of different size and bandwidth.[1] $D^*$ is calculated from the measured responsivity and dark current, $D^* = R(AB)^{1/2}/(2eI_d)^{1/2}$, where $R$ is the responsivity, $A$ is the device surface area (defined by the length of the channel and the overlap of the metal contact on the flake), $B$ is the bandwidth, and $I_d$ is the dark current. This constitutes an upper bound on the detectivity assuming the noise to be limited by shot noise. In Figure 4c we plot the calculated $D^*$ using the highest measured responsivity for each gate voltage at the peak response of 640 nm. Well below the high power threshold voltage (< -20 V) we use a bandwidth of 30 Hz estimated from the fall times in the photoconduction regime and above the threshold voltage we use a bandwidth of 0.1 Hz estimated from the fall time of the slow trap states. Due to the ultra-high responsivities measured in these devices the $D^*$ reaches $(3.3 \pm 0.8) \cdot 10^{13}$ Jones at a gate voltage of 20 V which is two orders of magnitude larger than shot noise limited values for photodetectors using MoS$_2$[7] and an order of magnitude larger than similar multilayer, two terminal In$_2$Se$_3$ photodetectors.[23]

**Conclusion:**

In summary, we have characterized multi-layer In$_2$Se$_3$ phototransistors in dark conditions and under illumination. We find that the dominate photocurrent generation mechanism can be tuned with the back gate from fast photoconduction in the OFF state to high gain photogating in the ON state. In the ON state, the surface oxide and long lived hole traps allow for ultra-high gain ($R \sim 10^5$ A/W) at a low bias voltage of 50 mV. Finally, for direct comparison with other nanostructured photodetectors, we calculate an inferred detectivity of $3 \cdot 10^{13}$ Jones making In$_2$Se$_3$ a promising material for photodetection applications.

**Methods:**
Fabrication of FET devices: In$_2$Se$_3$ flakes were exfoliated onto Si/SiO$_2$ (285 nm) substrates and contacts are patterned using standard e-beam lithography, thin-film metal deposition (5 nm Ti/ 60 nm Au), and lift off in hot acetone. The highly doped Si substrate is used as a back gate electrode. We define length of the channel as the distance between adjacent contacts and the width as the overlap distance of the electrode and flake. Due to the incomplete contact of the electrode and the flake we have included a 25% error in the width of the main text device. These errors propagate to the calculations of mobility, responsivity, and detectivity. Subsequent devices were created by directly transferring flakes onto pre-patterned contacts (5 nm Ti/ 30 nm Au) using an all dry transfer method reported in Ref. [38] to fabricate cleaner devices not previously exposed to e-beam resists. Three devices were studied in detail and two are presented in the main text and the supplement.

Optoelectronic measurements are performed in a *Lakeshore Cryogenics* probestation at room temperature in vacuum (<10$^{-5}$ mbar). Eight diode pumped solid state lasers are operated in continuous wave mode (CNI Lasers). The intensity is modulated by a mechanical chopper. The light is coupled into a multimode optical fiber through a parabolic mirror. At the end of the optical fiber, another identical parabolic mirror collimates the light exiting the fiber. The beam is then directed into the probe station's zoom lens system and then inside the sample space. The beam spot size on the sample has a 200 μm diameter for all wavelengths.

**Supporting Information:**
Supporting Information is available online. Raman and optical absorption characterizations of In$_2$Se$_3$ flakes, Optical image, AFM height scan, and gate hysteresis plots for main text device, Photocurrent data for contacts 1-2 and 2-3, Optoelectronic characteristics for another device.

**Acknowledgements:**
This work was supported by the Dutch organization for Fundamental Research on Matter (FOM) and from the Ministry of Education, Culture and Science (OCW). A.C-G. acknowledges financial support by the European Union through the FP7-Marie Curie Project PIEF-GA-2011-300802 ('STRENGTHNANO') and by the Fundacion BBVA through the fellowship 'I Convocatoria de Ayudas Fundacion BBVA a Investigadores, Innovadores y Creadores Culturales' (Semiconductores Ultradelgados: haoa la optoelectronica flexible).

**Figures:**

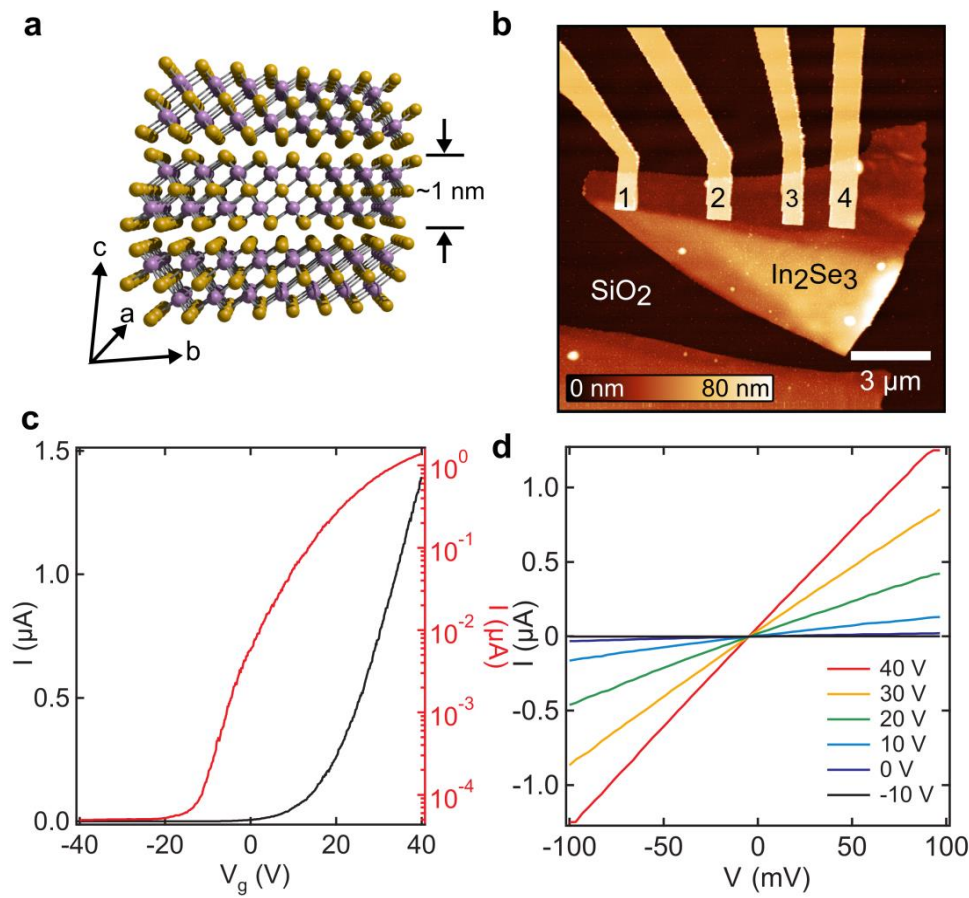

**Figure 1.** (a) Crystal structure of In$_2$Se$_3$. (b) AFM image of a representative FET device. (c) Transfer curve for the device in (b) using contacts 3 and 4. The black curve is plotted on a linear axis and red curve on a log axis. (d) *I-V* curves for different back-gate voltages measured between contacts 3 and 4.

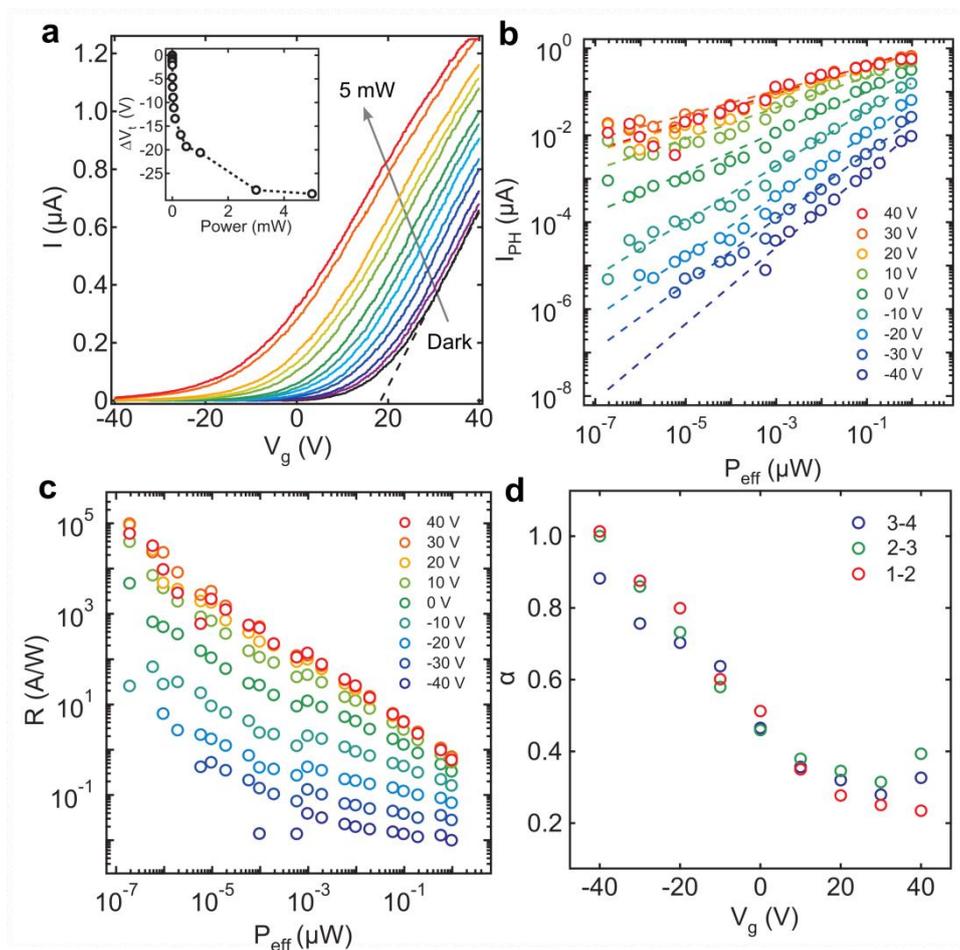

**Figure 2.** (a) Transfer curves for different illumination power of a 640 nm laser. The laser powers increase in series increments of 1, 3, 5 from 1 nW to 5 mW. Inset shows the shift of the threshold voltage for increasing powers. (b) Photocurrent vs. laser power for different back-gate voltages. (c) Responsivity calculated from the measured photocurrent in panel (b). (d) Exponent (α) extracted from panel (b) for each gate voltage.

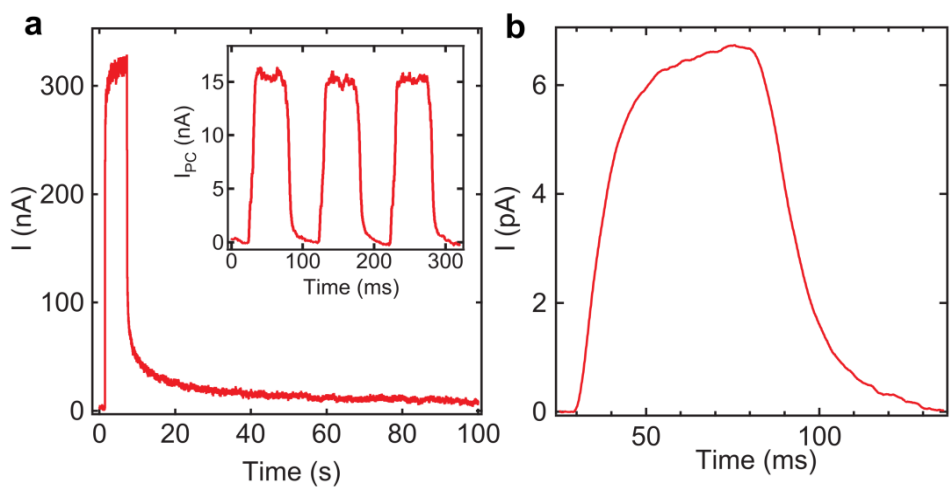

**Figure 3.** (a) Drain current measured as a function of time at $V_g = 0$ V. The rise a fall times are determined by blocking the laser (5 mW) with the chopper. Inset show the fast response that is present on top of the higher gain, slow response (at 1 mW power). Modulation frequency of the chopper is 10 Hz. (b) Single (averaged) illumination cycle (1 mW) at a gate voltage of -40 V. The modulation frequency of the chopper is 10 Hz.

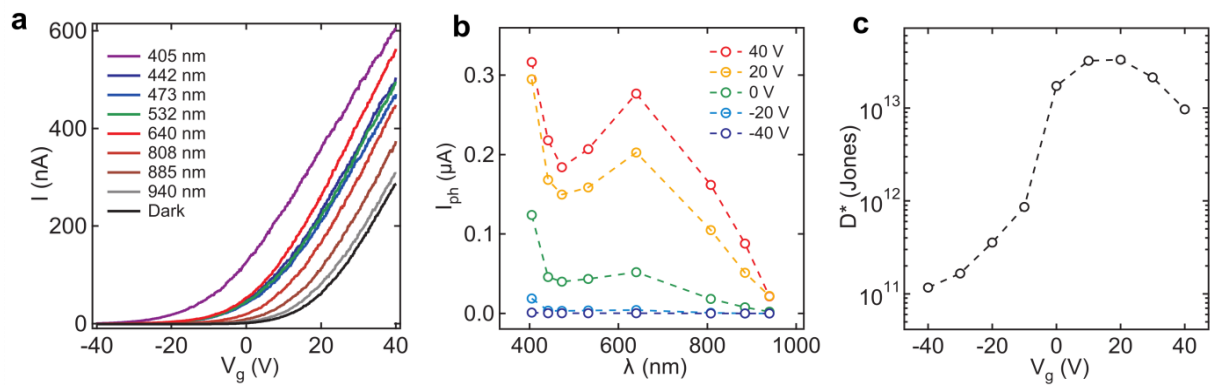

**Figure 4**. (a) Transfer curves for different illumination wavelengths. (b) Photocurrent extracted from (a) for different back-gate voltages. (c) Inferred detectivity ($D^*$) calculated from the maximum responsivities measured in Figure 2(c) and using the bandwidths estimated from Figure 3.

Table 1. Comparison with multilayer photodetectors having similar thicknesses

| Material | $V$ (V) | $V_g$ (V) | Responsivity (A/W) | Response Time (ms) | Reference |
|---|---|---|---|---|---|
| Multilayer In2Se3 | 0.05 | 40 | 98000 | 9000 | This work |
| Multilayer GaTe | 5 | 0 | 10000 | 6 | 16 |
| Multilayer TiS3 | 1 | -40 | 2900 | 4 | 36 |
| Multilayer In2Se3 | 5 | 0 | 390 | 18 | 10 |
| Multilayer MoSe2 | 20 | 0 | 97 | 30 | 39 |
| Multilayer InSe | 10 | 80 | 160 | 4000 | 40 |
| Multilayer InSe | 3 | 0 | 0.035 | 0.5 | 41 |

**REFERNCES:**

1. Buscema, M.; Island, J. O.; Groenendijk, D. J.; Blanter, S. I.; Steele, G. A.; van der Zant, H. S.; Castellanos-Gomez, A. *Chemical Society Reviews* **2015,** 44, (11), 3691-3718.
2. Xia, F.; Mueller, T.; Golizadeh-Mojarad, R.; Freitag, M.; Lin, Y.-m.; Tsang, J.; Perebeinos, V.; Avouris, P. *Nano Letters* **2009,** 9, (3), 1039-1044.
3. Xia, F.; Mueller, T.; Lin, Y.-m.; Valdes-Garcia, A.; Avouris, P. *Nature Nanotechnology* **2009,** 4, (12), 839-843.
4. Yin, Z.; Li, H.; Li, H.; Jiang, L.; Shi, Y.; Sun, Y.; Lu, G.; Zhang, Q.; Chen, X.; Zhang, H. *ACS Nano* **2011,** 6, (1), 74-80.
5. Alkis, S.; Oztas, T.; Aygün, L.; Bozkurt, F.; Okyay, A.; Ortaç, B. *Optics Express* **2012,** 20, (19), 21815-21820.
6. Choi, W.; Cho, M. Y.; Konar, A.; Lee, J. H.; Cha, G.-B.; Hong, S. C.; Kim, S.; Kim, J.; Jena, D.; Joo, J.; Kim, S. *Advanced Materials* **2012**, n/a-n/a.
7. Choi, W.; Cho, M. Y.; Konar, A.; Lee, J. H.; Cha, G. B.; Hong, S. C.; Kim, S.; Kim, J.; Jena, D.; Joo, J. *Advanced Materials* **2012,** 24, (43), 5832-5836.
8. Marcio Fontana, T. D., Anthony K. Boyd, Mohamed Rinzan, Amy Y. Liu, Makarand Paranjape, Paola Barbara. *arXiv:1206.6125 [cond-mat.mtrl-sci]* **2012**.
9. Yin, Z.; Li, H.; Jiang, L.; Shi, Y.; Sun, Y.; Lu, G.; Zhang, Q.; Chen, X.; Zhang, H. *ACS Nano* **2012,** 6, (1), 74-80.
10. Wang, Q. H.; Kalantar-Zadeh, K.; Kis, A.; Coleman, J. N.; Strano, M. S. *Nature Nanotechnology* **2012,** 7, (11), 699-712.
11. Lopez-Sanchez, O.; Lembke, D.; Kayci, M.; Radenovic, A.; Kis, A. *Nature Nanotechnology* **2013,** 8, (7), 497-501.
12. Mukherjee, B.; Cai, Y.; Tan, H. R.; Feng, Y. P.; Tok, E. S.; Sow, C. H. *ACS applied materials & interfaces* **2013,** 5, (19), 9594-9604.
13. Island, J. O.; Buscema, M.; Barawi, M.; Clamagirand, J. M.; Ares, J. R.; Sánchez, C.; Ferrer, I. J.; Steele, G. A.; van der Zant, H. S.; Castellanos‐Gomez, A. *Advanced Optical Materials* **2014,** 2, (7), 641-645.
14. Buscema, M.; Barkelid, M.; Zwiller, V.; van der Zant, H. S.; Steele, G. A.; Castellanos-Gomez, A. *Nano Letters* **2013,** 13, (2), 358-363.
15. Wu, C.-C.; Jariwala, D.; Sangwan, V. K.; Marks, T. J.; Hersam, M. C.; Lauhon, L. J. *The Journal of Physical Chemistry Letters* **2013,** 4, (15), 2508-2513.
16. Furchi, M. M.; Polyushkin, D. K.; Pospischil, A.; Mueller, T. *Nano Letters* **2014,** 14, (11), 6165-6170.
17. Ho, C.-H.; Lin, C.-H.; Wang, Y.-P.; Chen, Y.-C.; Chen, S.-H.; Huang, Y.-S. *ACS applied materials & interfaces* **2013,** 5, (6), 2269-2277.
18. Zhai, T.; Fang, X.; Liao, M.; Xu, X.; Li, L.; Liu, B.; Koide, Y.; Ma, Y.; Yao, J.; Bando, Y. *Acs Nano* **2010,** 4, (3), 1596-1602.
19. Li, Q.; Li, Y.; Gao, J.; Wang, S.; Sun, X. *Applied Physics Letters* **2011,** 99, (24), 243105.
20. Kwon, S. H.; Ahn, B. T.; Kim, S. K.; Yoon, K. H.; Song, J. *Thin Solid Films* **1998,** 323, (1), 265-269.
21. Peng, H.; Schoen, D. T.; Meister, S.; Zhang, X. F.; Cui, Y. *Journal of the American Chemical Society* **2007,** 129, (1), 34-35.
22. Liu, F.; Shimotani, H.; Shang, H.; Kanagasekaran, T.; Zolyomi, V.; Drummond, N.; Fal'ko, V. I.; Tanigaki, K. *ACS Nano* **2014,** 8, (1), 752-760.

23. Jacobs-Gedrim, R. B.; Shanmugam, M.; Jain, N.; Durcan, C. A.; Murphy, M. T.; Murray, T. M.; Matyi, R. J.; Moore, R. L.; Yu, B. *ACS Nano* **2013,** 8, (1), 514-521.
24. Marsillac, S.; Combot-Marie, A.; Bernede, J.; Conan, A. *Thin Solid Films* **1996,** 288, (1), 14-20.
25. Yu, X.; Hou, T.; Sun, X.; Li, Y. *Solid State Communications* **2013,** 162, 28-33.
26. Julien, C.; Hatzikraniotis, E.; Kambas, K. *physica status solidi (a)* **1986,** 97, (2), 579-585.
27. Luan, S.; Neudeck, G. W. *Journal of Applied Physics* **1992,** 72, (2), 766-772.
28. Weis, M.; Lin, J.; Taguchi, D.; Manaka, T.; Iwamoto, M. *Applied Physics Letters* **2010,** 97, (26), 263304.
29. Cui, J.; Wang, L.; Du, Z.; Ying, P.; Deng, Y. *Journal of Materials Chemistry C* **2015,** 3, (35), 9069-9075.
30. Sze, S. M.; Ng, K. K., *Physics of semiconductor devices*. John Wiley & Sons: 2006.
31. Kang, H.-S.; Choi, C.-S.; Choi, W.-Y.; Kim, D.-H.; Seo, K.-S. *Applied Physics Letters* **2004,** 84, (19), 3780-3782.
32. Takanashi, Y.; Takahata, K.; Muramoto, Y. *Electron Devices, IEEE Transactions on* **1999,** 46, (12), 2271-2277.
33. Zhang, W.; Huang, J. K.; Chen, C. H.; Chang, Y. H.; Cheng, Y. J.; Li, L. J. *Advanced Materials* **2013,** 25, (25), 3456-3461.
34. Soci, C.; Zhang, A.; Xiang, B.; Dayeh, S. A.; Aplin, D.; Park, J.; Bao, X.; Lo, Y.-H.; Wang, D. *Nano Letters* **2007,** 7, (4), 1003-1009.
35. González-Posada, F.; Songmuang, R.; Den Hertog, M.; Monroy, E. *Nano Letters* **2011,** 12, (1), 172-176.
36. Konstantatos, G.; Clifford, J.; Levina, L.; Sargent, E. H. *Nature photonics* **2007,** 1, (9), 531-534.
37. Zhang, E.; Jin, Y.; Yuan, X.; Wang, W.; Zhang, C.; Tang, L.; Liu, S.; Zhou, P.; Hu, W.; Xiu, F. *Advanced Functional Materials* **2015**.
38. Castellanos-Gomez, A.; Buscema, M.; Molenaar, R.; Singh, V.; Janssen, L.; van der Zant, H. S.; Steele, G. A. *2D Materials* **2014,** 1, (1), 011002.
39. Abderrahmane, A.; Ko, P.; Thu, T.; Ishizawa, S.; Takamura, T.; Sandhu, A. *Nanotechnology* **2014,** 25, (36), 365202.
40. Tamalampudi, S. R.; Lu, Y.-Y.; Kumar U, R.; Sankar, R.; Liao, C.-D.; Moorthy B, K.; Cheng, C.-H.; Chou, F. C.; Chen, Y.-T. *Nano Letters* **2014,** 14, (5), 2800-2806.
41. Lei, S.; Ge, L.; Najmaei, S.; George, A.; Kappera, R.; Lou, J.; Chhowalla, M.; Yamaguchi, H.; Gupta, G.; Vajtai, R. *ACS Nano* **2014,** 8, (2), 1263-1272.

Supporting Information

# Gate controlled photocurrent generation mechanisms in high-gain In$_2$Se$_3$ phototransistors

*J. O. Island\*, S. I. Blanter\*, M. Buscema, H. S.J. van der Zant, A. Castellanos-Gomez*

**Supporting Information Contents**

1. **Raman and optical absorption of an In$_2$Se$_3$ flake**
   **Figure S1:** Raman

   **Figure S2:** Absorption

2. **Optical image, AFM height scan and gate hysteresis of main text device**
   **Figure S3:** Optical image, AFM height scan and gate sweeps

3. **Photocurrent data for contacts 1-2 and 2-3**
   **Figure S4:** Photocurrent vs. P$_{eff}$

4. **Optoelectronic characteristics for another device**

### 1. Raman and optical absorption of an In$_2$Se$_3$ flake

Alpha phase In$_2$Se$_3$ flakes (from 2dsemiconductors.com) are characterized using Raman spectroscopy and absorption spectroscopy. Raman spectroscopy differentiates the alpha and gamma phases from the remaining three phases but due to the similarities in the Raman spectrum of the alpha and gamma phases, we have also performed absorption spectroscopy to measure the optical band gap. The alpha phase has a reported band gap of 1.453 eV[1] and the gamma phase has a reported band gap of 1.812 eV[2].

In Figure S1 we show the optical image and Raman spectra of an exfoliated flake on a Si/SiO$_2$ substrate. Raman peaks are observed at 181 cm$^{-1}$ and 200 cm$^{-1}$ corresponding to the A$_1$ modes of α-phase In$_2$Se$_3$.[3]

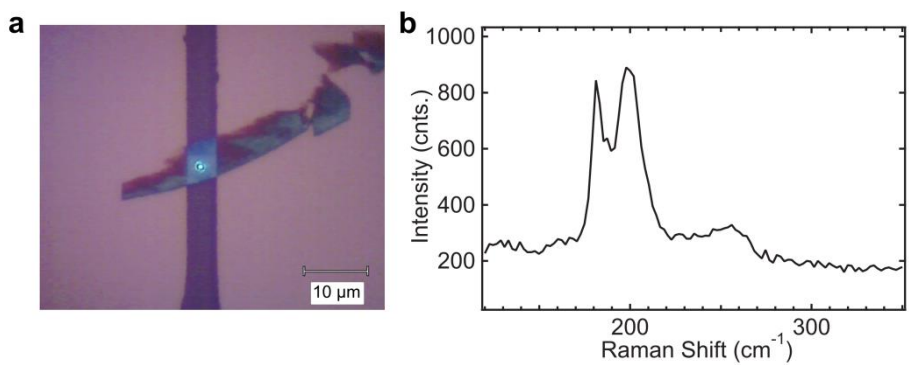

**Figure S1:** (a) Optical image of an exfoliated flake. The laser spot can be seen at the center of the flake at the position where the spectra was taken. (b) Raman spectra showing peaks at the location of the A$_1$ modes for α-In$_2$Se$_3$

Furthermore, we determine the bandgap of the material by measuring its absorption spectra by transferring an exfoliated flake onto the core of a multimode optical fiber using an all-dry viscoelastic stamping method.[4] Figure S2 shows optical images of the transferred flake and the absorption spectra. To determine the bandgap, we plot $(AE)^2$ where A is the measured absorption and E is the energy, and extrapolate the gap from a linear fit to the spectra. Using this method we determine a bandgap of 1.43 eV in close agreement with reported values of the α-phase of In$_2$Se$_3$ (1.453 eV).[1]

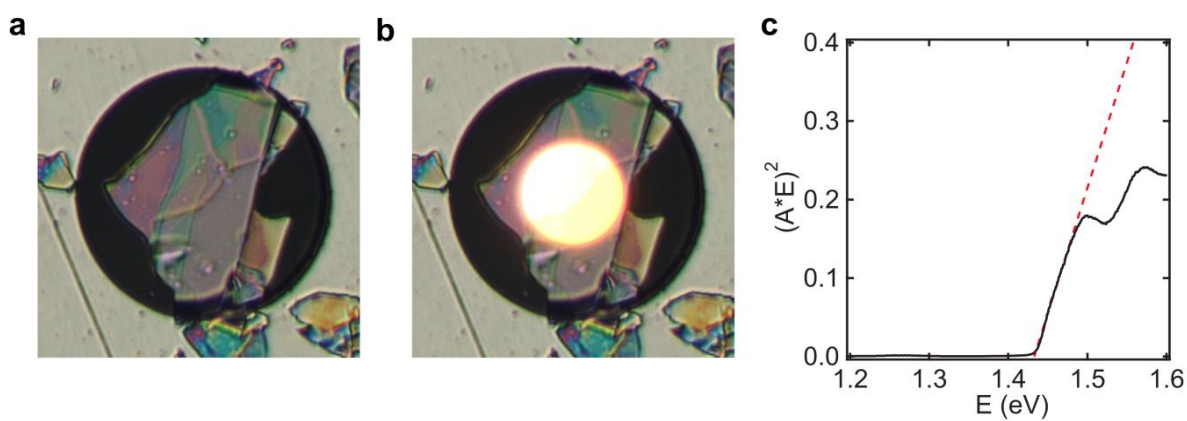

**Figure S2:** (a) Optical image of an exfoliated In$_2$Se$_3$ flake that has been transferred onto the core of a multimode optical fiber. (b) Lighted region shows the core of the multimode optical fiber. (c) Absorption spectra of the flake used to determine the bandgap.

2. **Optical image, AFM height scan and gate hysteresis of main text device**

Optical and AFM images of the main text device are shown in Figure S3. Optically, flakes of ~ 10 nm can be found by a bluish color under a white-light optical microscope (Figure S3a). The height scan in Figure S3c shows that this flake is about 14 nm thick, corresponding to roughly 15 layers. In Figure S3d we show the transfer curves for contacts 1-2 for a forward sweep (black curve) and reverse sweep (red curve). The hysteresis in the transfer characteristics is a common sign of charge trapping at the flake surface and interface.[5, 6] From the hysteresis curve ($\Delta V_t$ = 11 V) we estimate a trap density of $C\Delta V_t$ ~ $10^{12}$ cm$^{-2}$, where C is the capacitance to the back gate.

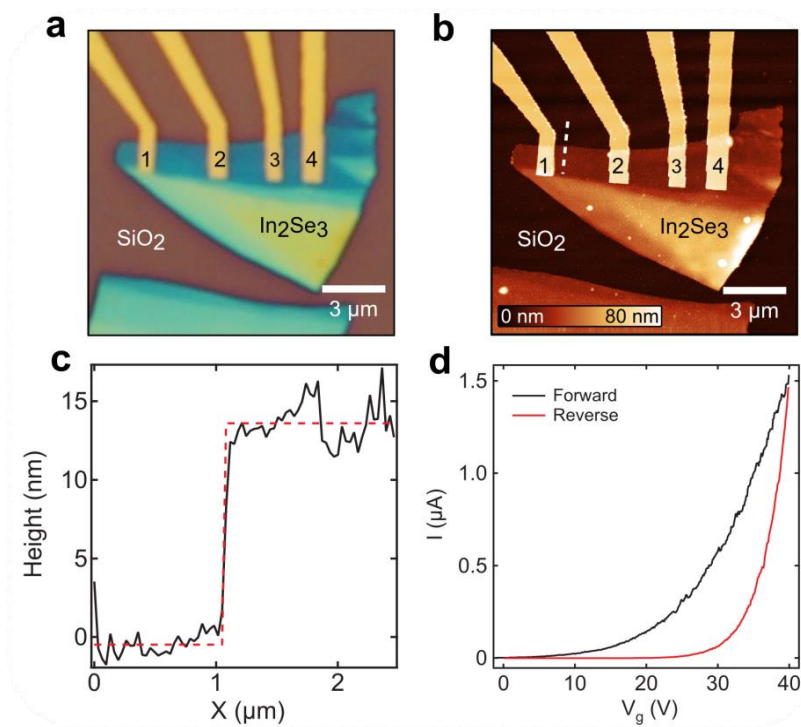

**Figure S3**: (a) Optical image of the main text device. (b) AFM image of the same device. (c) Height scan taken at the location of the white dotted line in panel (b). (d) Transfer curve for a forward (black curve) and reverse sweep (red curve).

3. **Photocurrent data for contact 1-2 and 2-3**

In Figure S4 we show the measured photocurrent as a function of effective incident laser power for different back gate voltages. Linear fits are made to the higher effective powers (~$10^{-4}$ µW to ~1 µW) to extract the alpha exponent plotted in the main text Figure 2d.

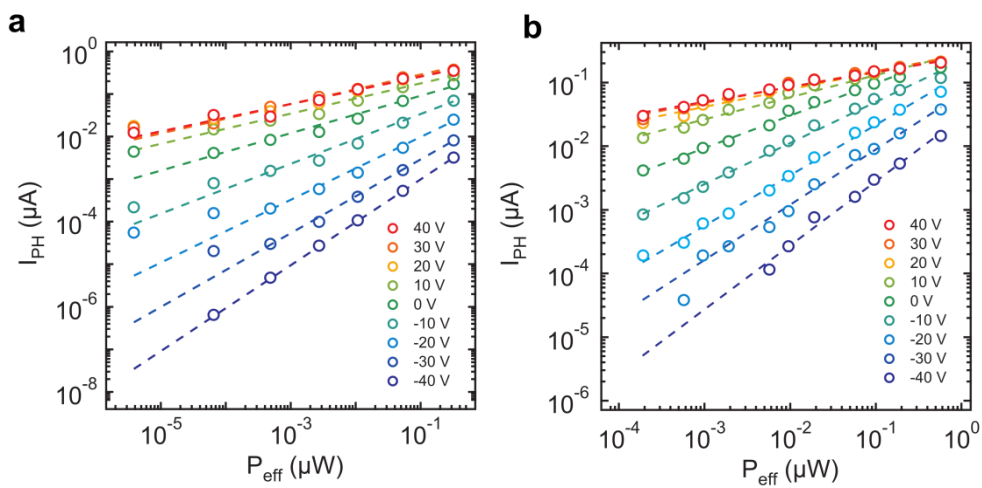

**Figure S4:** (a) Measured photocurrent ($I_{PH}$) as a function of effective power ($P_{eff}$) for different gate voltages for contacts 2-3 of the main text device. (b) same measurements for contacts 1-2.

4. **Optoelectronic characteristics for another device**

In an attempt to study further the effect of residues from fabrication processes (resists) we have also fabricated devices by simply transferring flakes onto pre-patterned electrodes. Figure S5 shows an AFM and dark FET characteristics for a second In$_2$Se$_3$ device. In order to reduce contamination from fabrication, this device has been fabricated by simply transferring a flake onto pre-patterned gold electrodes. Comparing with the main text device, the mobility is lower (10$^{-2}$ cm$^2$/Vs, estimated FET mobility from transfer curve) which we attribute to poor contact between the electrodes and transferred flake. ON/OFF ratios are also lower due to the increased thickness.

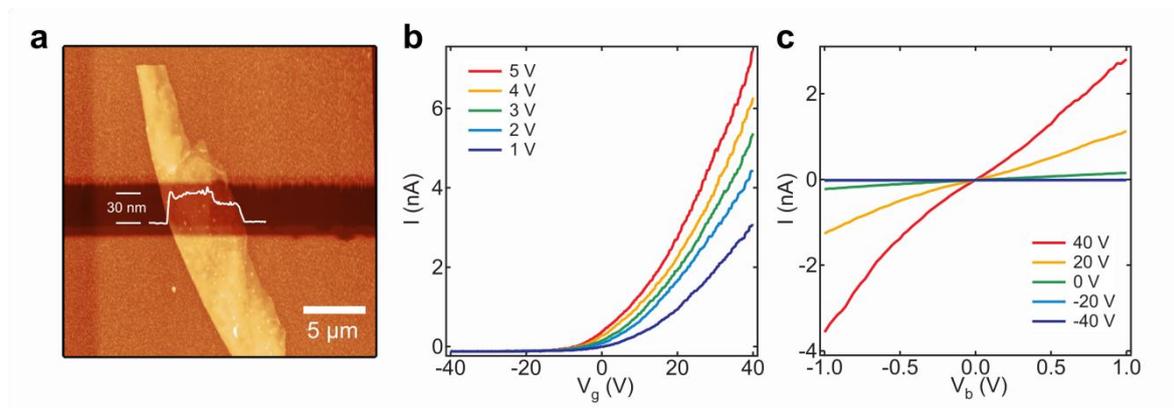

**Figure S5:** (a) AFM image of a transferred In$_2$Se$_3$ flake onto pre-patterned electrodes. (b) Transfer curves at bias voltages from 1 V to 5 V for the device in (a). (c) Source-drain (I-V) curves at back gate voltages from -10 V to 40 V.

We additionally characterized the device under illumination and observed the same characteristics found in the main text device. Transfer curves are presented in Figure S6a for increasing laser powers. The threshold voltage steadily shifts and saturates at higher powers (Figure S6a inset). The responsivities are more than an order of magnitude lower (Figure S6c) but this could also be a consequence of poor contact to the flake resulting in lower mobilities. The photocurrent response can be similarly tuned with back gate voltage going from a linear dependence on incident power ($\alpha \approx 1$) in the OFF state (Vg = -40 V) to ($\alpha \approx 0.5$) in the ON state (Vg = 40 V) (see Figure S6d).

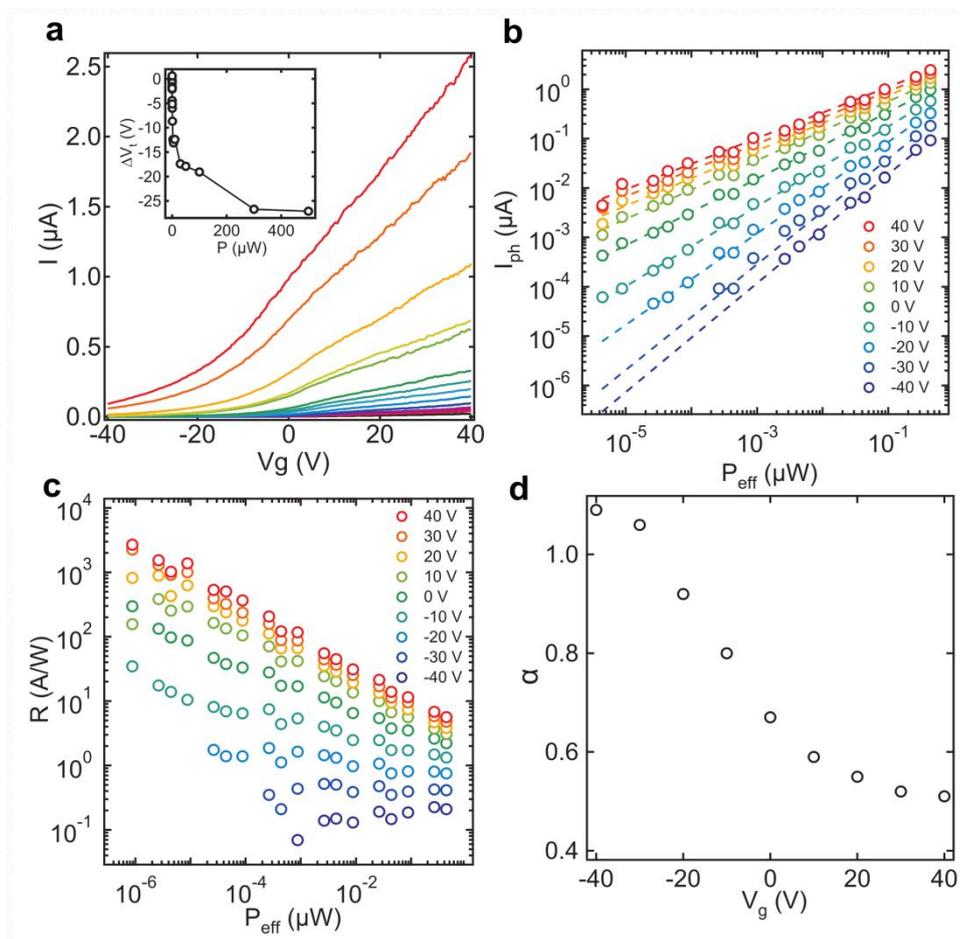

**Figure S6:** (a) Transfer curves for different illumination power of a 640 nm laser. The laser powers increase in series increments of 1, 3, 5 from 1 nW to 500 µW. Inset shows the shift of the threshold voltage for increasing powers. (b) Photocurrent vs. laser power for different back-gate voltages. (c) Responsivity calculated from the measured photocurrent in panel (b). (d) Exponent (α) extracted from panel (b) for each gate voltage.

In Figure S7 we present the time and spectral response for this device. Figure S7a shows three on/off cycles for the device at a chopper frequency of 19 Hz. The OFF state photoconductive effect response is slightly faster for this device (fall time ≈ 2 ms). Figure S7b shows the spectral response. Again we measure a peak response around a laser wavelength of 640 nm coinciding with the energy gap of the natural oxide.

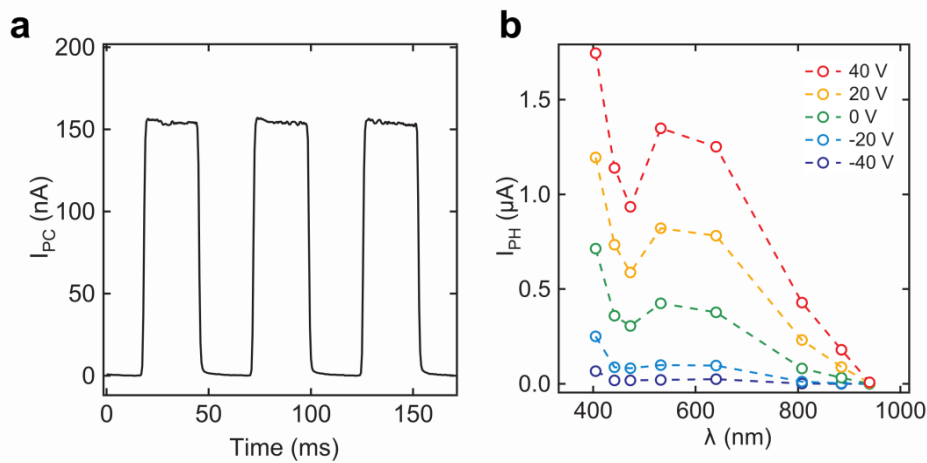

**Figure S7:** (a) Drain current measured as a function of time at Vg = -40 V. The rise a fall times are determined by blocking the laser (1 mW) with the chopper (b) Photocurrent extracted from transfer curves for different laser wavelengths.

Overall the characteristics are similar to the main text device but the responsivities are an order of magnitude lower $< 10^4$ A/W and the response times are an order of magnitude faster (2 ms). This is an indication that the devices are cleaner with less low energy traps. Although the oxide is still present in the spectral response, the absence of fabrication residues helps speed up response at the cost of gain.

REFERENCES:
1. Ho, C.-H.; Lin, C.-H.; Wang, Y.-P.; Chen, Y.-C.; Chen, S.-H.; Huang, Y.-S. *ACS applied materials & interfaces* **2013,** 5, (6), 2269-2277.
2. Julien, C.; Chevy, A.; Siapkas, D. *physica status solidi (a)* **1990,** 118, (2), 553-559.
3. Lewandowska, R.; Bacewicz, R.; Filipowicz, J.; Paszkowicz, W. *Materials Research Bulletin* **2001,** 36, (15), 2577-2583.
4. Castellanos-Gomez, A.; Buscema, M.; van der Zant, H. S.; Steele, G. A. *arXiv preprint arXiv:1311.4829* **2013**.
5. Kim, W.; Javey, A.; Vermesh, O.; Wang, Q.; Li, Y.; Dai, H. *Nano Letters* **2003,** 3, (2), 193-198.
6. Late, D. J.; Liu, B.; Matte, H. R.; Dravid, V. P.; Rao, C. *ACS Nano* **2012,** 6, (6), 5635-5641.